\documentclass[11pt]{iopart}

\usepackage{graphicx}
\usepackage{amssymb}

\usepackage{float}
\usepackage{tabularx}
\usepackage{microtype}
\usepackage{hyperref}
\usepackage{xcolor}
\usepackage[T1]{fontenc} 
\usepackage{lmodern}

\usepackage{cite}

\begin{document}



\title[]{Divertor shaping with neutral baffling as a solution to the tokamak power exhaust challenge}



\author{K. Verhaegh$^{1}$, J.R. Harrison$^1$, D. Moulton$^1$, B. Lipschultz$^2$, N. Lonigro$^{2,1}$, N. Osborne$^{3,1}$, P. Ryan$^1$, C. Theiler$^4$, T. Wijkamp$^{5,6}$, D. Brida$^7$, C. Cowley$^{2,1}$, G. Derks$^{6,5}$, R. Doyle$^8$, F. Federici$^{9}$, B. Kool$^{6,5}$, O. Février$^4$, A. Hakola$^{10}$, S. Henderson$^1$, H. Reimerdes$^4$, A.J. Thornton$^1$, N. Vianello$^{11}$, M. Wischmeier$^7$, L. Xiang$^1$ and the EUROfusion Tokamak Exploitation Team$^{*}$ and the MAST Upgrade Team$^{**}$}
\address{$^1$ United Kingdom Atomic Energy Authority, Culham, United Kingdom} 
\address{$^2$ York Plasma Institute, University of York, United Kingdom}
\address{$^3$ University of Liverpool, Liverpool, United Kingdom}
\address{$^4$ Swiss Plasma Centre, \'{E}cole Polytechnique F\'{e}\'{e}rale de Lausanne, Lausanne, Switzerland}
\address{$^5$ Eindhoven University of Technology, Eindhoven, The Netherlands}
\address{$^6$ Dutch Institute for Fundamental Energy Research DIFFER, Eindhoven, The Netherlands}
\address{$^7$ Max Planck Institute for Plasma Physics, Garching, Germany}
\address{$^8$ Dublin City University, Dublin, Ireland}
\address{$^9$ Oak Ridge National Laboratory, Oak Ridge, United States}
\address{$^{10}$ VTT Technical Research Centre of Finland, Espoo, Finland}
\address{$^{11}$ Consorzio RFX, Padova, Italy}
\address{$^{*}$ See the author list of “Overview of the EUROfusion Tokamak Exploitation programme in support of ITER and DEMO” by E. Joffrin Nuclear Fusion 2024 10.10788/1741-4326/ad2be4.}
\address{$^{**}$ See the author list of “Overview of physics results from MAST Upgrade towards
	core-pedestal-exhaust integration” by J.R. Harrison et al. to be published in Nuclear Fusion Special Issue: Overview and Summary Papers from the 29th Fusion Energy Conference (London, UK, 16-21 October 2023).}
\ead{kevin.verhaegh@ukaea.uk}

\begin{abstract}
	Exhausting power from the hot fusion core to the plasma-facing components is one of the biggest challenges in fusion energy. The MAST Upgrade tokamak uniquely integrates strong containment of neutrals within the exhaust area (divertor) with extreme divertor shaping capability. By systematically altering the divertor shape, this study shows the strongest evidence to date that long-legged divertors with a high magnetic field gradient (total flux expansion) deliver key power exhaust benefits without adversely impacting the hot fusion core. These benefits are already achieved with relatively modest geometry adjustments that are more feasible to integrate in reactor designs. Benefits include reduced target heat loads and improved access to, and stability of, a neutral gas buffer that 'shields' the target and enhances power exhaust (detachment). Analysis and model comparisons shows these benefits are obtained by combining multiple shaping aspects: long-legged divertors have expanded plasma-neutral interaction volume that drive reductions in particle and power loads, while total flux expansion enhances detachment access and stability. Containing the neutrals in the exhaust area with physical structures further augments these shaping benefits. These results demonstrate strategic variation in the divertor geometry and magnetic topology is a potential solution to one of fusion's biggest challenges: power exhaust.
\end{abstract}


\section{Introduction}
\label{ch:introduction}

Sustainable nuclear fusion is one of the most promising solutions for the world's energy challenges, offering an essentially limitless and clean energy source. However, one of the critical hurdles in developing viable fusion reactors is efficiently managing its power exhaust: removing heat and particles from the hot fusing core while reducing surface heat fluxes to sufficiently low levels to prevent damaging the reactor's components \cite{Wenninger2014,Zohm2021}. Using novel experiments, analysis and model results from the MAST Upgrade tokamak, this study not only demonstrates that innovative shaping of the power exhaust region can solve this critical challenge, but also explains the physics and synergy between combining different power exhaust shaping strategies.

\begin{figure}[ht]
	\centering
	\includegraphics[width=\linewidth]{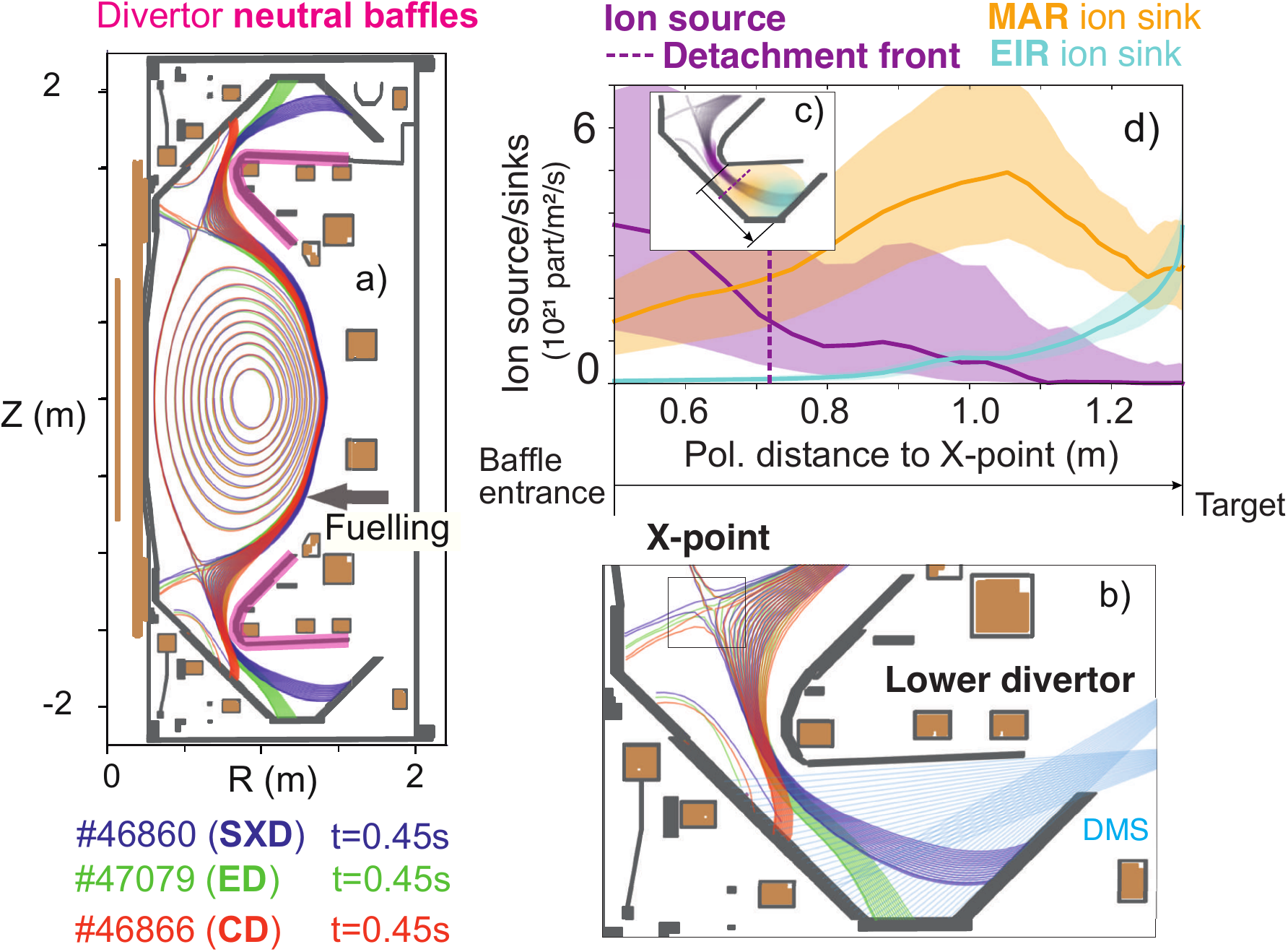}
	\caption{Overview of MAST-U plasma shapes and coil positions, combined with an overview of the plasma processes in the MAST-U divertor. (a) Overview of the magnetic geometry for the Super-X Divertor (SXD), Elongated Divertor (ED) and Conventional Divertor (CD), together with the fuelling and divertor baffle location. (b) Lower divertor with diagnostic coverage of the Divertor Monitoring Spectrometer (DMS) \cite{Verhaegh2023,Verhaegh2023b}. (c) Schematic illustration of the characteristic processes in a detached MAST-U Super-X divertor. (d) 1D profile of the, spectroscopically, line-integrated inferences (ions/$m^2$/s) of the divertor ion sources (magenta) and Molecular Activated Recombination (MAR, orange) and Electron-Ion Recombination (EIR, cyan) ion sinks as function of poloidal distance from the X-point to the target for the Super-X divertor, obtained from advanced spectroscopic analysis \cite{Verhaegh2021} for \# 46860 at 45 \% Greenwald fraction. The detachment (or ionisation) front position is indicated with a dotted magenta line in both (c) and (d).}
	\label{fig:MagGeom}
\end{figure}

In magnetic confinement fusion, such as tokamaks and stellarators \cite{Effenberg2019}, the hot fusion core plasma is confined within nested magnetic field lines ('closed flux tubes'). Heat and particles are expelled from the core into the edge region, where they follow the 'open flux tubes' forming the Scrape-Off Layer (\autoref{fig:MagGeom}a). Coils enable altering the magnetic topology of these open flux tubes to create a magnetic null point ('X-point'), which diverts heat and particle fluxes to a dedicated region called the 'divertor' (\autoref{fig:MagGeom}b). Since the power exhaust is carried by charged particles following the flux surfaces, the narrow width of the SOL results in extreme heat fluxes (150 $MW/m^2$ for the DEMO reactor design \cite{Wenninger2014,Zohm2021}) due to the narrow plasma wetted area, far exceeding engineering limits (5-10 MW/$m^2$ \cite{Wenninger2014,Zohm2021}) if unmitigated.

To reduce target heat loads, the power must be spread over a larger area. This is first achieved by injecting radiating impurity gasses or hydrogen fuel to cool the divertor plasma and convert the heat carried by charged particles into heat carried by photons (radiation) that do not follow the magnetic field lines and thus dissipate the power volumetrically \cite{Stangeby2000}. However, cooling the divertor plasma increases the ion target fluxes and associated power loading from surface recombination, limiting the total possible power reduction to a factor $\sim 4$ \cite{Verhaegh2021}. 'Divertor detachment' is a process that reduces the ion target flux, enabling further power dissipation and order-of-magnitude target heat flux reductions \cite{Lipschultz1998,Krasheninnikov2017,Verhaegh2021,Stangeby2000}. At electron temperatures of $\leq \sim 3-5$ eV, the ionising plasma 'detaches' from the target, forming a neutral buffer below the ionising plasma or 'detachment'/ionisation front (\autoref{fig:MagGeom}(c,d)). Plasma-atom/molecule interactions, within that buffer, cause simultaneous power, particle (e.g. ion), and momentum losses, that collectively drive detachment (see \nameref{Methods} section). Recombining the ions into neutral atoms, through ion sinks like Molecular Activated Recombination (MAR) and Electron-Ion Recombination (EIR) (\autoref{fig:MagGeom}(c,d)), plays a critical role in detachment \cite{Lipschultz1998,Verhaegh2021,Verhaegh2023,Krasheninnikov2017}. One drawback of detachment is that it can be highly sensitive to changes in core power, impurity seeding and fuelling. A high sensitivity could more easily result in a loss of detachment, damaging the reactor walls, or in the detached region reaching the hot fusing core, resulting in a radiative collapse of the plasma \cite{Goetz1996} that can catastrophically damage a reactor \cite{Maris2024}. Therefore, we refer to reducing the detachment sensitivity as an increased detachment stability for simplicity.

Despite advances in understanding and maximising the mitigation of heat/particle fluxes through plasma detachment, maintaining core performance while effectively exhausting power remains a major challenge and key uncertainty for future reactors \cite{Zohm2021}. Compact reactor designs like STEP \cite{Hudoba2023,Osawa2023}, SPARC \cite{Kuang2020}, ST-F1/E1 \cite{Windridge2020} and ARC \cite{Labombard1995,Wigram2019}, aiming to accelerate the pathway to fusion energy \cite{Labombard2015,Windridge2020}, face even larger power exhaust challenges. Innovative power exhaust solutions are thus required for compact fusion reactors \cite{Hudoba2023,Kuang2020,Wigram2019} and, as risk mitigation, for DEMO and beyond \cite{Xiang2021,Kembleton2022,Militello2021}. This includes (combinations of) liquid metal targets \cite{Eden2017,Lore2022}, high impurity injection to induce X-point radiators \cite{Stroth2022,Pan2022,Lunt2023}, and Alternative Divertor Configurations (ADCs) \cite{Theiler2017,Ryutov2015,Reimerdes2017}. ADCs use coils to optimise the divertor magnetic topology to reduce heat loads \cite{Theiler2017} whilst maintaining a hot fusion core; increase the range of core conditions for which detachment can be achieved \cite{Theiler2017} and improve the stability of detachment \cite{Lipschultz2016} (see \nameref{Methods} section).

One promising ADC approach combines long-legged divertors \cite{Umansky2020,Reimerdes2017}, achieved by increasing the distance between the X-point and the target to increase the power dissipation volume, with using the divertor magnetic topology to spread the power over a larger target area: poloidal and total flux expansion. Poloidal flux expansion increases the distance between poloidal magnetic flux tubes (poloidal flux expansion $F_x = \frac{B_\theta^{u} B_\phi^t}{B_\theta^t B_\phi^u}$ see \nameref{Methods}), whereas total flux expansion increases the magnetic field gradient by increasing the target radius (total flux expansion $F_R = \frac{B_t}{B_{xpt}}$). ADCs can be further optimised by containing the neutral particles within the divertor chamber using baffle plates (\autoref{fig:MagGeom}a), boosting plasma-neutral interactions and preventing neutrals escaping to the core where they cool the fusion core plasma, thus enhancing core-edge compatibility \cite{Reimerdes2021,Umansky2020}.

MAST Upgrade is the UK's national fusion experiment, newly built to tackle fusion's power exhaust challenge. MAST-U's design uniquely integrates strong neutral baffling, long-legged divertors ((poloidal) divertor leg length / major radius $> 1$), and high total flux expansion (up to 2.5). In contrast, conventional divertor solutions (on JET, Asdex-Upgrade \cite{Eich2017}) have short-legged divertors (divertor leg length / major radius $< 0.1$), negligible total flux expansion ($F_R \sim 1$) and no neutral baffles. The spherical ('apple-shaped') nature of MAST-U enables $F_R$ variations over a much larger range (1-2.5) than possible in conventional ('doughnut-shaped') tokamaks with flexible shaping, such as TCV (1-1.6) \cite{Theiler2017}. Preliminary MAST-U results under low power (Ohmic) conditions ($P_{SOL}=0.4$ MW) demonstrate the benefits of the 'Super-X Divertor' (SXD) \cite{Valanju2009,Fil2019submitted}, which has the highest $F_R$ achievable, over the conventional divertor ($F_R = 1.2$) \cite{Verhaegh2023,Verhaegh2023c,Moulton2023}; consistent with simulations \cite{Moulton2023}. 

This work shows the key experimental results of exploring novel divertor solutions on MAST-Upgrade. Since there is a continuum of ADC solutions, this work studies the impact of varying total flux expansion and divertor leg length instead of focusing only on the extreme Super-X topology, using plasmas with higher power (1.5-1.7 MW Neutral Beam Injection (NBI) heating, $P_{SOL} = 1.2$ MW). This not only provides the strongest experimental evidence to date for the benefits of ADCs by combining total flux expansion, divertor leg length and neutral baffling; but also shows these benefits are maintained at more moderate divertor shaping (lower $f_R$ and shorter leg lengths than the maximum values). These power exhaust benefits are obtained without any adverse core impact and include target heat flux reductions; improved access to, and stability of, plasma detachment; as well as improved core-edge compatibility. Since engineering complexity is a critical hurdle for integrating ADCs in reactors, the finding that their benefits can be maintained at more moderate shaping is of key importance for fusion reactor engineering and design \cite{Militello2021,Kembleton2022}, advancing the path for using ADCs to achieve sustainable fusion energy. 

Low temperature plasma physics and plasma chemistry are central to plasma detachment. In this work we unravel these processes as a function of divertor shape for the first time through advanced analysis techniques \cite{Verhaegh2023} and model comparisons. This shows novel insights into how different shaping aspects work together to achieve the observed benefits. The divertor poloidal leg length/volume results in additional power/particle losses without impacting the plasma upstream (section \ref{ch:PowerPartBal}). Total flux expansion improves access to, and stability of, detachment (section \ref{ch:ReduModel}). Strong neutral baffling enables total flux expansion benefits and augments plasma-neutral interactions, maximising the benefits of divertor poloidal leg length (section \ref{ch:TCV_comparison}). This provides unprecedented understanding of how divertor shaping can improve power exhaust in agreement with model predictions (sections \ref{ch:ReduModel}, \ref{ch:ShapeScan}), further advancing the path for using ADCs to achieve sustainable fusion energy.

\section{Results}

By systematically comparing three divertor geometries: the Conventional Divertor (CD); Elongated Divertor (ED) and Super-X Divertor (SXD) (\autoref{fig:MagGeom}, divertor shape parameters shown in Table 1), we obtain five benefits of combined total flux expansion, poloidal leg length and divertor neutral baffling. 

\begin{enumerate}
	\item Improved access to detachment: detachment occurs at lower core density.
	\item Increased operational regime for detached divertor operation: the range of core density and powers at which the divertor detached is larger.
	\item Improved detachment stability: the sensitivity of detachment to changes in core density is reduced.
	\item Reduced target heat fluxes and power loads.
	\item Improved power exhaust without adverse core impact: core performance in detached conditions is improved.
\end{enumerate}

For each divertor configuration, the evolution of their power exhaust and detachment properties are diagnosed as the core electron density is gradually increased and the divertor conditions grow colder, whilst other parameters are held as constant as possible. For ease of reference and for comparison against literature, the line-averaged core electron density (obtained from interferometry) is expressed as a fraction to the maximum core density, Greenwald, limit \cite{Greenwald1988} which depends on the plasma current and tokamak size.

\begin{table}[ht]
	\begin{tabular}{lllllll}
		Discharge & $R_t$ (m)   & $F_x$ & $F_R$ & L (m)   & $L_{pol}$ (m) & Description                                \\
		46860     & 1.45        & 9     & 2.3           & 19      & 1.3            & Super-X Divertor (SXD)                     \\
		47079     & 1.11        & 6     & 1.7           & 17      & 1.1            & Elongated Divertor (ED)                    \\
		46762     & 0.79        & 3.3   & 1.2           & 13      & 0.64           & Conventional Divertor (CD)                 \\
		46895     & 0.81 - 1.39 & 4 - 6 & 1.2 - 2.2     & 13 - 19 & 0.65 - 1.3     & CD -\textgreater ED -\textgreater SXD scan
	\end{tabular}
	\label{tab:ShapeParam}
	\caption{Summary of the magnetic divertor shape parameters for the discharges discussed in this work, featuring target radius ($R_t$), poloidal flux expansion ($F_x$), total flux expansion ($F_R$), connection length from the upstream midplane to the target ($L$) and poloidal leg length from the X-point to the target ($L_{pol}$). Discharge \# 46895 keeps the core density and power constant as the divertor topology is changed over the range of the first three discharges.}.
\end{table}

\subsection{Benefits of combined strong neutral baffling, total flux expansion and divertor leg length}
\label{ch:Benefits}

\begin{figure}[ht]
	\centering
	\includegraphics[width=\linewidth]{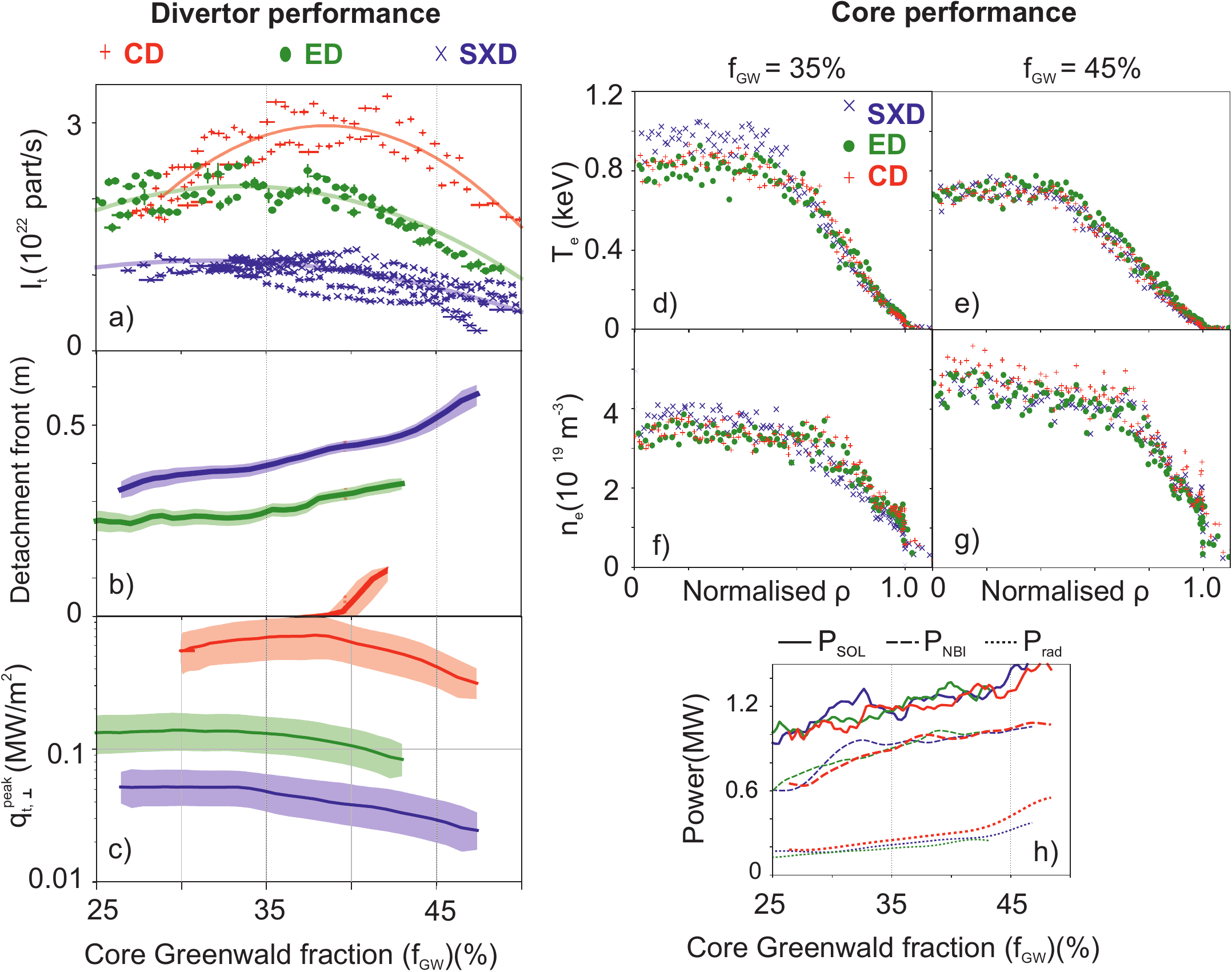}
	\caption{Comparison of divertor (a,b,c) and core (d,e,f,g,h) performance as function of core Greenwald fraction ($f_{GW}$ in \%) for the CD (red), ED (green) and SXD (blue). Divertor parameters: (a) Integrated ion target flux (with polynomial fits), (b) detachment (ionisation) front position as poloidal distance to the target (m), (c) estimated perpendicular target heat load combining Langmuir probe and spectroscopy measurements (see \nameref{Methods}) \cite{Verhaegh2023,Verhaegh2023b}. Core parameters: (d-g) core electron temperatures and densities at two different core Greenwald fractions (corresponding to vertical dotted lines in (a,b,c,h)), (h) $P_{SOL}$ (solid lines) deduced from the following contributors: NBI absorption (TRANSP, dashed lines); Ohmic heating (EFIT, not shown); changes to stored energy (EFIT, not shown) and core radiative losses (bolometry, dotted lines).}
	\label{fig:DivCoreOverview}
\end{figure}

\emph{The longer legged, totally flux expanded, divertors have improved access to detachment} at lower core plasma densities. In the CD, the integrated target particle fluxes (\autoref{fig:DivCoreOverview}a) increase as function of core density, indicative of an attached divertor plasma, up to a core Greenwald fraction of $f_{GW} \approx 40 \%$. At this point, both the particle flux at the target decreases and the ionisation front detaches from the target (\autoref{fig:DivCoreOverview}b), indicative of the onset of detachment. In contrast, the ED and SXD are detached throughout the scanned core density range: the particle flux does not increase with increasing density whilst the ionisation front remains detached. Since there is no difference in the density limit achievable between the different geometries, \emph{the operational window} (in terms of the core density range for which the discharge is detached) \emph{for detached operation is increased for the long-legged divertors} compared to the CD.

\emph{The longer-legged, totally flux expanded, divertors have a higher detachment stability} to quasi-steady-state changes in core parameters, qualitatively consistent with reduced (steady-state) models (section \ref{ch:DLS_model} \cite{Lipschultz2016}). The sensitivity of the detachment front to changes in core density, i.e, the slope of the detachment front position (\autoref{fig:DivCoreOverview}(b)) is a factor 5 steeper for the CD ($f_{GW} \approx 40 \%$), compared to the ED and SXD at this core density: the detachment front is much more sensitive to changes in core density for the CD. These \emph{steady-state} results indicate an inherent stabilisation, akin to a shock absorber, of the detachment front for the ED and SXD, in contrast to the CD where the ionisation region moves with minimal core changes out of the divertor chamber after detachment, increasing core and X-point radiation (\autoref{fig:DivCoreOverview}(h)). Although the reduced detachment sensitivity and increased operational window of detachment are related, they are not identical. Using the analogy of a shock absorber, a reduced detachment sensitivity corresponds to a stronger damping, whereas the wider detached operational regime increases the displacement the spring can undergo before the elastic limit is exceeded: both work in unison \emph{enhancing detachment stability} for long-legged, totally flux expanded divertors. 

These benefits extend to \emph{dynamic} variations in fuelling \cite{Kool2024} and heating perturbations \cite{Verhaegh2023c}: indicating an inherent lesser response of the detachment location to fuelling/heating transients and improved detachment control \cite{Kool2024} for the ED and SXD. In contrast to our quasi-steady-state experiments, \cite{Kool2024} features sinusoidal core fuelling oscillations with more than 7 times faster fuelling changes to which the divertor responds dynamically. 

\emph{The longer-legged, totally flux expanded, divertors result in larger heat flux reductions than expected}. Based on the magnetic geometry \cite{Theiler2017}, a reduction in perpendicular heat flux by $\sim 5.8 \times $ and $\sim 2.1 \times$ for the SXD and ED is expected, compared to the CD, mainly due to increased poloidal ($F_x \sim 3.0 \times$ (SXD) and $\sim 1.6 \times$ (ED)) and total ($F_R \sim 2.0 \times$ (SXD) and $\sim 1.4 \times$ (ED)) flux expansion. However, a much larger reduction in target heat flux is observed: $\sim 18.5 \times$ and $\sim 7 \times$ for the SXD and ED, compared to the CD (\autoref{fig:DivCoreOverview} c). The longer-legged divertors result in \emph{additional heat flux dissipation} through volumetric and/or radial/cross-field transport by a factor $\sim 3.2 \times$, qualitatively consistent with both SOLPS-ITER simulations (section \ref{ch:ShapeScan}) \cite{Moulton2023} and volumetric power loss estimates (section \ref{ch:PowerPartBal}).

\emph{The longer-legged, totally flux expanded, divertors enable divertor detachment without adverse core impact}, in contrast to the CD - which needs high densities to detach ($f_{GW}>40\%$). The core densities, temperatures and $P_{SOL}$ are similar for the CD, ED and SXD (\autoref{fig:DivCoreOverview}(d-h)). These results indicate a strong decoupling between the divertor shape and the obtained core conditions, even when the outer target is detached. 

\subsection{What drives the physics of long-legged, totally flux expanded, strongly baffled divertors ?}

After having shown the benefits of strongly baffled, long-legged, totally flux-expanded, divertors, we will explore why these divertor configurations have a superior exhaust performance using spectroscopic analysis \cite{Verhaegh2021b}, reduced models and simulation comparisons. This shows poloidal leg length, total flux expansion and strong neutral baffling strengthen each other's impact and all work together to achieve the observed benefits. The additional leg volume in the SXD and ED, compared to the CD, results in their superior power dissipation and drives the reduction of the ion target flux during detachment through ion sinks, whereas total flux expansion drives reductions in detachment onset and improves detachment front stability. Neutral baffling augments plasma-neutral interactions, strengthening the benefits of poloidal leg length, and enables total flux expansion benefits by preventing neutral leakage to the core.

\subsubsection{Additional volume long-legged divertors drives power and particle losses}
\label{ch:PowerPartBal}

\begin{figure}[ht]
	\centering
	\includegraphics[width=\linewidth]{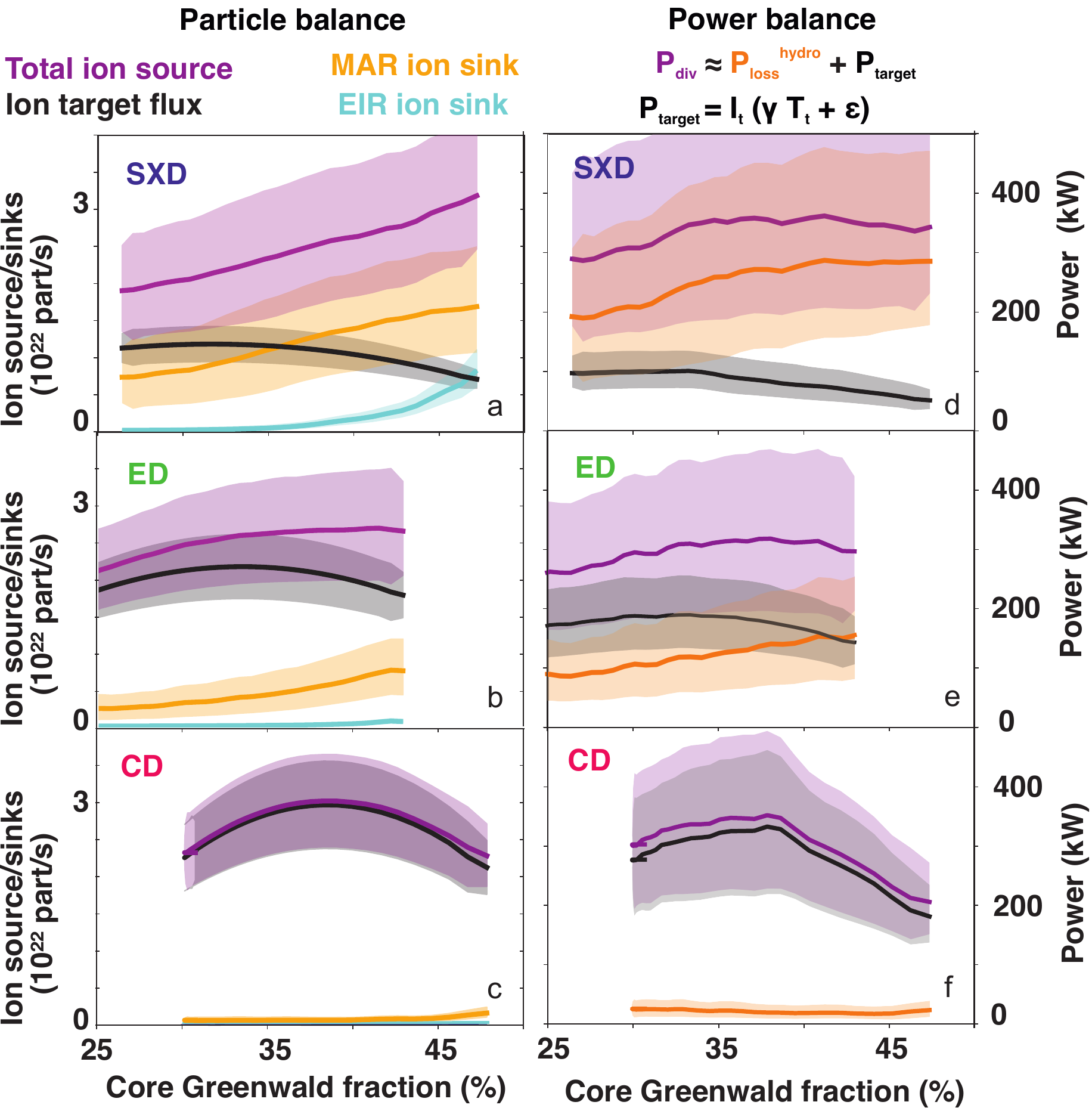}
	\caption{Particle (a-c) and power (d-f) balance comparisons between different divertor shapes as function of core Greenwald fraction. (a-c) Particle balance showing the ion target flux (lower outer divertor), total ionisation source, MAR and EIR ion sinks (both ion sinks are integrated over the divertor chamber) for the SXD (a), ED (b) and CD (c). (d-f) Power balance showing hydrogenic power losses $P_{loss}^{hydro}$ (integrated over the divertor chamber), target power deposition $P_{target}$ (obtained from spectrocopically inferred temperatures and Langmuir probe particle fluxes) and estimated power flow into the divertor chamber ($P_{div} \approx P_{loss}^{hydro} + P_{target}$) assuming that the divertor chamber power losses are dominantly hydrogenic, in agreement with imaging bolometry measurements \cite{Verhaegh2023c}. Under the assumption that the lower and upper divertors are similar (consistent with Langmuir probe results \cite{Verhaegh2023c}), $P_{div}$, $P_{loss}^{hydro}$ and $P_{target}$ have been multiplied by two to obtain integrated values of the upper and lower outer divertors.}
	\label{fig:PartPowerBalSXD_ED_CD}
\end{figure}

Particle balance analysis shows \emph{increased ion sinks reduce the target fluxes} in the SXD and ED compared to the CD (\ref{fig:PartPowerBalSXD_ED_CD}(a,b,c)), whereas the total ion source is the same within uncertainties for all three different geometries. Ion sinks are significant in both the SXD and the ED from the start of those discharges, both through MAR as well as EIR (in the SXD). Our spectroscopic analysis reveals plasma conditions of $n_e = 2-4 \times 10^{19} m^{-3}$ and $T_e \approx 0.2$ eV in the region where EIR becomes observable ($f_{GW}>33\%$ in the SXD and $f_{GW}>40\%$ in the ED) \cite{Verhaegh2023c, Lonigro2023}. MAR only appears in the CD at the highest core densities after its ionisation front detaches from the target ($f_{GW} > 40 \%$), but its magnitude remains limited downstream the baffle.

The \emph{total ion source}, inferred through particle balance, is obtained by adding the ion target flux and the ion sinks observed in the divertor chamber (see \nameref{Methods} section). This consists out of the divertor chamber ion source and any net inflow of ions into the divertor chamber. Previous work \cite{Verhaegh2023c} showed the ion source is generated between the X-point and the ionisation front and up to 40\% of it is generated in the divertor chamber. X-point imaging shows the neutral baffling is effective in limiting the ion source upstream of the X-point for all three configurations, consistent with simulation results (Figure \ref{fig:xpt_imag_sim}, section \ref{ch:ShapeScan}).

Analogously to the particle flux reduction, it is the additional volumetric power dissipation in their divertor volume (\autoref{fig:PartPowerBalSXD_ED_CD} (d-f)) that drives the reduction in target power loads for the SXD and ED. The inferred power flowing into the divertor chamber is similar for all three geometries (\autoref{fig:PartPowerBalSXD_ED_CD} (d-f) for $f_{GW} < 40 \%$). As the inferred hydrogenic radiation is similar to the total measured radiation from an imaging bolometer (not shown) \cite{Verhaegh2023b}, the divertor chamber power losses mostly arise from hydrogenic processes. These hydrogenic power losses reduce $P_{target}$ by a factor $\sim \times 4$ (SXD) and $\sim \times 2$ (ED) compared to the CD; consistent with the target heat load reduction being larger than expected based on geometry (\autoref{fig:DivCoreOverview}(c)). A significant part of the ED and SXD hydrogenic power losses originate in the detached regime from Molecular Activated Dissociation (MAD). Elastic collisions between the plasma and the neutral cloud, which is neglected above, can further augment the power losses in the ED and SXD by up to 15\% of $P_{SOL}$ \cite{Osborne2024} according to the SOLPS-ITER simulations shown in section \ref{ch:ShapeScan}. This would occur in the detached region and depend on the plasma-neutral interaction volume downstream the ionisation front, which is increased for longer-legged divertors \cite{Osborne2024}. 

The ED and SXD maintain strong radiation in the divertor chamber. That is in contrast to the CD configuration where the radiation upstream of the divertor baffle lowers $P_{div}$ after the detachment onset ($f_{GW} > 40 \%$), consistent with the much higher detachment front location sensitivity to changes in the core density (\autoref{fig:DivCoreOverview}b). These results illustrate longer-legged divertors can 1) increase maximum (divertor) power dissipation; 2) maintain power losses away from the X-point towards the divertor target; and 3) further enhance power losses after the onset of detachment.

\begin{figure}[ht]
	\centering
	\includegraphics[width=\linewidth]{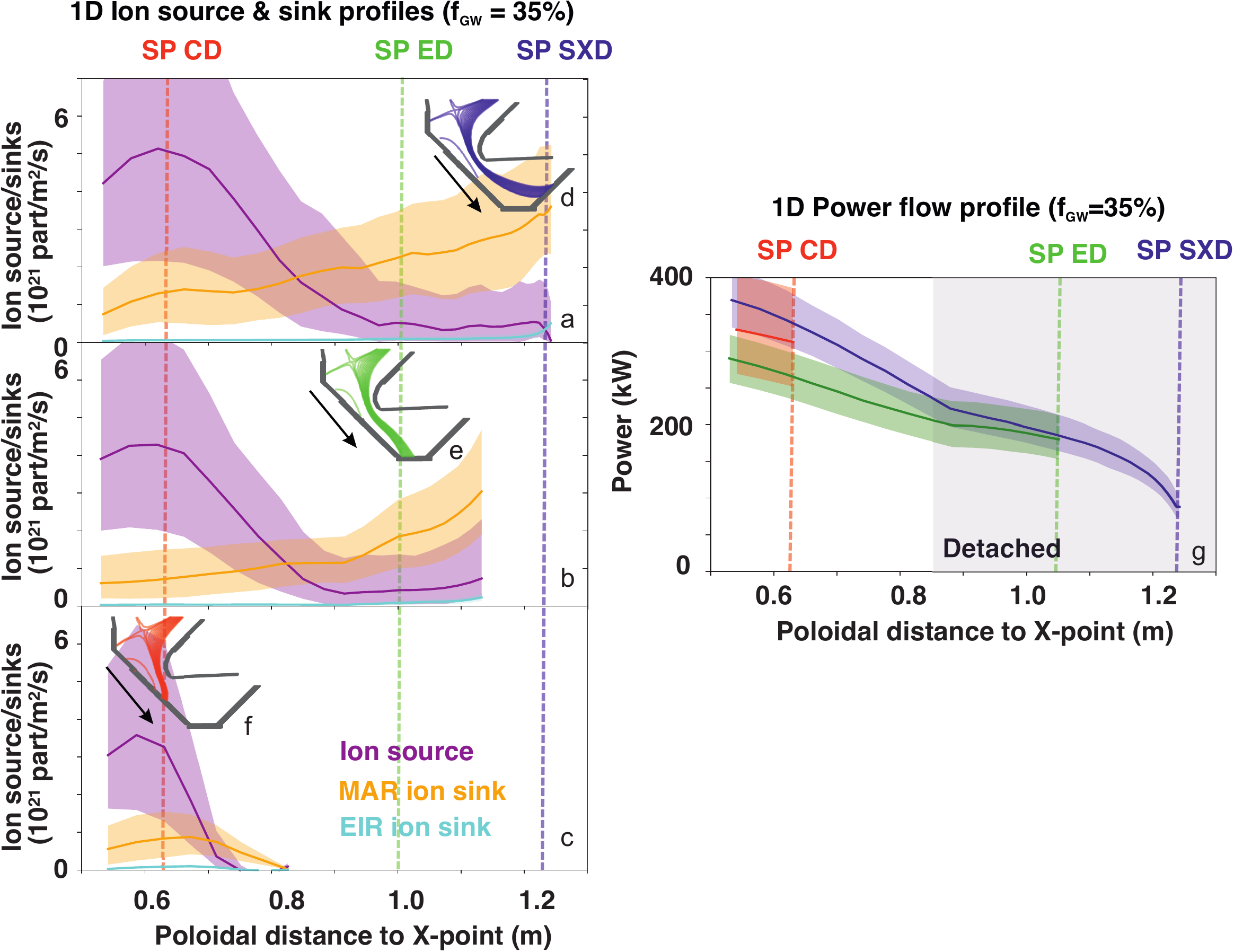}
	\caption{The additional divertor leg length/volume of the SXD and ED configurations drives additional power and particle losses, whereas the ion sources/sinks and power losses are similar at the same poloidal distance to the X-point, according to the their 1D profiles along the divertor leg. Spectroscopically inferred line-integrated ion sources and sinks (part./$m^2$/s) for the SXD (a), ED (b) and CD (c) at $f_{GW} = 35 \%$ as function of poloidal distance to the X-point. The coloured dotted lines indicate their respective strike point positions, indicated by their magnetic geometry (d-f). The 1D ion source/sink profiles (a,b,c) are extended downstream of their respective strike-points due to convolution of the radial-extent of the SOL/far-SOL with the spectroscopic lines-of-sight, where the plasma is colder than at the separatrix. (g) Power flow ($W$) towards the divertor targets as function of poloidal distance to the X-point from the divertor entrance to the target for the CD (red), ED (green) and SXD (blue) at $f_{GW} = 35 \%$, with a dotted line indicating their respective strike points. The part where the divertor leg is detached is indicated in grey. The power flow is inferred by subtracting from $P_{div}$ the cumulative sum of the hydrogenic power losses from upstream to the target.}
	\label{fig:PowerBalFlowProf}
\end{figure}

The impact of divertor shaping on power and particle exhaust is revealed when studying the 1D profiles of ion sources and sinks (a-c), as well as power flows (g), along the divertor leg \emph{as function of poloidal distance to the X-point} at a fixed core density ($f_{GW} = 35 \%$) (section \ref{fig:PowerBalFlowProf}). Both profiles are similar between the different geometries (up until the CD detachment onset) \emph{at the same poloidal distance to the X-point}: the plasma is thus predominantly altered in the extended region. The deeper detachment and lower power loads in the SXD and ED are brought on by interactions in the additional volume available downstream of the ionisation region when the divertor leg is extended.  Plasma-chemistry occurring in this region, resulting in MAR and MAD, plays a key role explaining the differences between the different divertor geometries.


\subsubsection{Total flux expansion drives detachment onset and sensitivity reduction}
\label{ch:ReduModel}

To gain further insights into the impact of divertor shaping on detachment, the experimental results are compared against the Detachment Location Sensitivity (DLS) analytical model \cite{Lipschultz2016,Cowley2022,Myatra2023a}. The DLS model predicts detachment occurs if the parameters driving detachment (in our case core density ($f_{GW}$) and power ($P_{SOL}$), lumped together as $C \propto \frac{n_{e,u} \sqrt{f_I}}{q_\parallel^{5/7}} \propto \frac{f_{GW}}{P_{SOL}^{5/7}}$ - see \nameref{Methods} section) reaches the detachment threshold, $C_t$.  $C_t$ is a function of the magnetic geometry: $C_t \propto \frac{1}{F_R} (\frac{B_{xpt}}{<B>})^{2/7} \frac{1}{L_\parallel^{2/7}}$, depending mostly on total flux expansion ($F_R$), connection length ($L_\parallel$, parallel to the field line) and the averaged magnetic field strength $<B>$. Total flux expansion and connection length both lower $C_t$, reducing the density (and impurity content - see \nameref{Methods} section) required for detachment at a fixed $P_{SOL}$.

The impact of total flux expansion on the detachment onset predicted by the DLS model is consistent with our observations (table \ref{tab:DLS_vs_experiment}). The DLS predicted benefits for the SXD and ED over the CD arise mostly from an increase in total flux expansion. Given the density at which the CD detaches, the DLS predicts that the SXD and ED are already detached at the lowest core density achieved, consistent with the experiment. To compare the SXD and ED against each other, the onset of EIR is used as a colder reference point for DLS comparisons, showing agreement within 10 \% of the experiment. Furthermore, the DLS model predicts that the detachment front position only depends on the magnetic field topology upstream of the detachment front (see \nameref{Methods} section), consistent with the observed invariance of the upstream parameters to the downstream divertor magnetic topology. 

\begin{table}[]
	\begin{tabular}{lllll}
		& \multicolumn{2}{l}{Detachment onset} & \multicolumn{2}{l}{EIR onset} \\ \hline
		& Model       & Experiment             & Model   & Experiment          \\ \hline
		SXD & $-55\%$       & $<-37\%$       & $0\%$     & Reference                 \\
		ED  & $-40\%$       & $<-37\%$       & $+36\%$   & $+27\%$               \\
		CD  & $0\%$         & Reference                    & $+106\%$  & $>+52\%$ \\ \hline
	\end{tabular}
	\caption{DLS reduced model predictions are in agreement with observations. Measured and DLS predicted relative differences (in core density) between the different divertor topologies for the detachment onset ($T_e = \sim 3-5$ eV) and the onset of EIR, serving as a colder reference point ($T_e \ll 1$ eV). The experimentally observed density at which the CD detaches ($f_{GW}^{ref,CD,detach}$) and at which the EIR occurs in the SXD ($f_{GW}^{ref,SXD,EIR}$) are used as reference densities. The percentages shown are the observed and DLS modelled relative density differences to the reference for detachment onset ($f_{GW}^{ED,SXD,detach}/f_{GW}^{ref,CD,detach}$) and EIR onset ($f_{GW}^{CD,ED,EIR}/f_{GW}^{ref,SXD,EIR}$). These DLS modelled differences only depend on magnetic topology, see \nameref{Methods} section. The density range obtainable in the experiment limits the differences in detachment and EIR onset that can be explored between the different topologies. Therefore, when $<$ (or $>$) is indicated, the detachment or EIR onset is not observed in the experiment and the relative difference is larger than indicated.}.
	\label{tab:DLS_vs_experiment}
\end{table}

Although only the magnetic geometry is used for obtaining DLS predictions on the differences between the SXD, ED and CD, it should be noted that the DLS model derivation assumes that: 1) the detachment front is infinitely thin; 2) all power dissipation is driven by impurity radiation. Both these assumptions are invalid for the MAST-U conditions shown. However, the agreement between the DLS model and the MAST-U results suggests that the impact of divertor topology on the detachment onset may be more generally applicable outside of impurity radiation dominant conditions. Further work is required generalising the DLS model for MAST-U like conditions, although the validity of these assumptions would increase in more reactor-relevant conditions.

Our power and particle balance analysis showed the \emph{additional volume} in the SXD and ED is critical to explain the reduction of \emph{power and particle loads during detachment}, whereas \emph{total flux expansion} drives 80\% of the improved \emph{access to detachment} according to DLS model comparisons, with the longer divertor leg length driving the remaining 20\%. The DLS model also predicts that total flux expansion reduces detachment sensitivity, qualitatively consistent with ED and SXD observations (benefit iii): the combination of increasing divertor leg length/volume and total flux expansion result in strong, synergistic, power exhaust benefits. 

\subsubsection{Neutral trapping enables exhaust benefits of total flux expansion and poloidal leg length}
\label{ch:TCV_comparison}

Although the poloidal divertor leg length downstream the baffle entrance is very different between the the different topologies, our experiments suggest that the baffling has a similarly strong impact on the CD (up until its detachment onset), ED and SXD. The observed and simulated (section \ref{ch:ShapeScan}) divertor neutral pressures are similar between the three geometries. Likewise, the neutral trapping, defined as the ratio between the ion source downstream the X-point to the total ion source \cite{Fil2019submitted}, is similar between the three divertor topologies in simulations (78 \% for the SXD and ED and 75 \% for the CD, respectively), despite the SXD and ED configurations being deeply detached (section \ref{ch:ShapeScan}, Figure \ref{fig:xpt_imag_sim}). In contrast, SOLPS-ITER simulations for a core density ramp of the (open, un-baffled) conventional TCV divertor indicated a neutral trapping of 32-45 \% in detachment onset conditions, decreasing during deeper detachment to 11 \% \cite{Verhaegh2023a,Fil2019submitted}. However, the increased detachment front sensitivity of the CD likely results in a reduction of neutral trapping after its detachment onset, diminishing the impact of neutral baffling on the CD after its detachment onset.

The benefits of long-legged divertors, total flux expansion, divertor shaping and neutral baffling have been individually studied on TCV \cite{Reimerdes2017,Theiler2017}; showing benefits of neutral baffling \cite{Reimerdes2021,Raj2022,Gorno2023,Sheikh2021} and long divertor leg lengths \cite{Reimerdes2017}. The benefits of total flux expansion (both in terms of detachment onset and front sensitivity/stability) were, however, much smaller than predicted by the DLS model \cite{Carpita2023,Fevrier2021}. Escape of neutrals from the divertor to the SOL upstream of the X-point can lead to strong plasma flows from the midplane to the target \cite{Cowley2024,Havlickova2015}, which diminishes the impact of total flux expansion on the detachment onset \cite{Carpita2023}. SOLEDGE2D-EIRENE simulations suggest that the neutral baffling on TCV may be insufficient to recover the full benefit of total flux expansion \cite{Meineri2023}, consistent with previous SOLPS-ITER simulations \cite{Fil2019submitted,Fil2017} which showed that the neutral trapping of the (open, unbaffled) TCV Super-X divertor was worse than that of the conventional divertor. This not only negates part of the benefits of total flux expansion due to an ion flow from upstream the X-point towards the target, but also reduces the benefit of the neutrals in power/particle dissipation between the Super-X divertor compared to the conventional divertor.

This difference between MAST-U and TCV, as well as the absence of strong flows from the midplane to the X-point in MAST-U simulations, suggests that strong neutral trapping, obtained by baffling on MAST-U, \emph{enables} the \emph{shaping} benefits of total flux expansion. Additionally, neutral baffling augments power/momentum/particle losses from plasma-neutral interactions such as MAR and MAD by containing the neutrals in the divertor chamber, amplifying the benefits of long-legged divertors (section \ref{ch:PowerPartBal}). Although neutral baffling is not required to obtain strong neutral trapping on high power devices (i.e., JET, AUG) near attached conditions, they would still be required to maintain high neutral trapping in cases where the ionisation source is significantly upstream of the target, motivating the STEP \cite{Henderson2024}, SPARC \cite{Kuang2020} and ARC \cite{Wigram2019} divertor designs. The MAST-U results suggest that neutral baffling can be placed downstream the X-point: it is important that the baffle structure is: 1) upstream of the intended location of the ionisation front; 2) sufficiently prohibits the escape of neutrals to the SOL.

\subsection{Strike point variations further confirm shaping benefits, consistent with exhaust simulations}
\label{ch:ShapeScan}

MAST-U \cite{Morris2018} uniquely integrates strong baffling and extreme divertor shaping, enabling strongly baffled novel divertor solutions. The DLS detachment onset predictions ($C_t$) varies by 110 \% between MAST-U CD and SXD. Comparatively, $C_t$ varies by 70 \% for a range of TCV divertor geometries (target radius scan, poloidal flux expansion scan, as well as for the X-point target divertor) \cite{Theiler2017,Reimerdes2017}. By integrating strong baffling, long-legged divertors and total flux expansion, MAST-U enables retrieving the full benefit of its extreme shaping capability on the detachment onset and stability (factor $5 \times$ reduced sensitivity for ED \& SXD vs CD, \autoref{fig:DivCoreOverview}(b)), as well as power and particle exhaust capability. We will now investigate the results of a single discharge under constant constant core density (30 \% Greenwald fraction, $n_e^{sep} \approx 0.8\times 10^{18} m^{-3}$) and power ($P_{SOL} \approx 1.0$ MW), where $C_t$ was altered by 110 \% by slowly sweeping the outer strike point from a CD to an ED to a SXD geometry. No significant differences are observed, at the same strike point position and core density, during this strike point scan at constant core density with the density ramps presented previously.

\begin{figure}[ht]
	\centering
	\includegraphics[width=\linewidth]{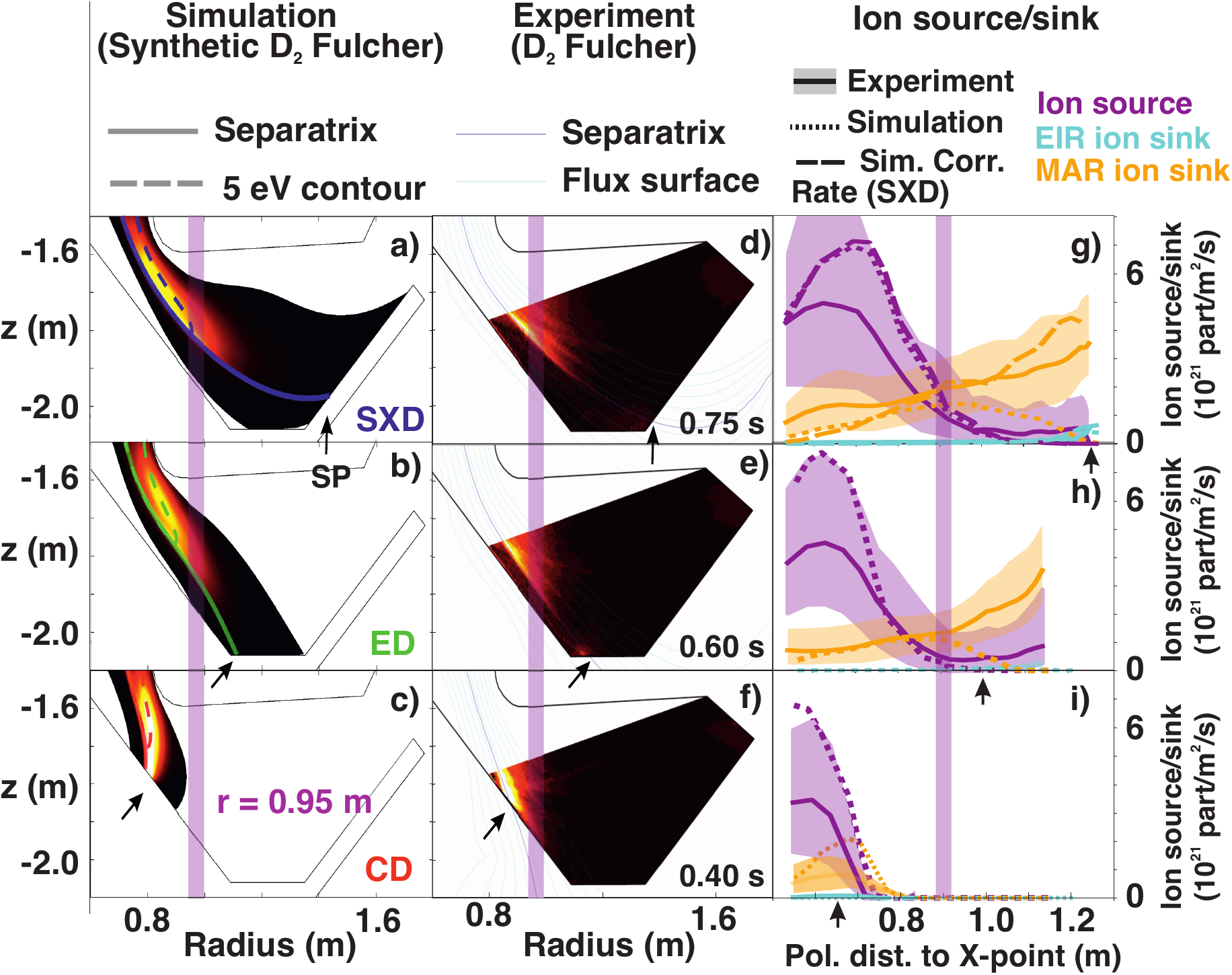}
	\caption{Experimental data shows the ionisation region location is unaltered after detachment during a strike point scan, in quantitative agreement with SOLPS-ITER simulations. (a-c) Synthetic $D_2$ Fulcher emission from SOLPS-ITER simulations for the SXD (a), ED (b) and CD (c), overlaid with 5 eV contours (dotted lines) and the separatrix (solid line). (d-f) Experimentally measured $D_2$ Fulcher band emission (595-605 nm) for a strike point scan, moving from CD to SXD at constant density and power, obtained through inverting Multi-Wavelength-Imaging (MWI) imaging data for \# 46895 \cite{Wijkamp2023}. (g-i) 1D ion sources and sinks,  obtained from spectroscopic analysis integrated along the spectroscopic lines of sight (\autoref{fig:MagGeom}b) (part./$m^2$/s), compared against synthetic diagnostic results from SOLPS-ITER simulations (dotted lines). For the SXD (g) two SOLPS-ITER simulation results are shown: one with default rates and one with corrected molecular charge exchange ($D_2 + D^+ \rightarrow D_2^+ + D$) rates ('Sim. Corr. Rate'), obtained from \cite{Verhaegh2023c}, which increases MAR. To guide the eye, a magenta vertical line has been added at a radius of 0.95 m and an arrow has been added at the strike point location (a-i). }
	\label{fig:FulcherSOLPSDetachEvolution}
\end{figure}

The results in figure \ref{fig:DivCoreOverview}(b) suggested that the ionisation front position, once detached from the target, is invariant to the magnetic topology downstream of it: its location depends only on the upstream magnetic topology. This is confirmed by the strike point sweep discharge, where the core conditions (not shown) remain mostly unchanged, with a 5-10\% increase in upstream and core $T_e$ when transitioning from CD to SXD (potentially due to the longer connection length \cite{Stangeby2000}). After the  $D_2$ Fulcher band emission front, which is a proxy for the ionisation front \cite{Verhaegh2023,Wijkamp2023,Osborne2023}, detaches from the target (target radius $\approx 0.95$ m), it remains close to this radial position as the strike point is swept further and further outwards and both total flux expansion and poloidal leg length is increased (\autoref{fig:FulcherSOLPSDetachEvolution} (d-f)). This implies that the ionisation front position (for this $P_{SOL}$ and $f_{GW}$) remains at a constant poloidal distance to the X-point as the divertor leg length is further increased.

\begin{figure}
	\includegraphics[width=\linewidth]{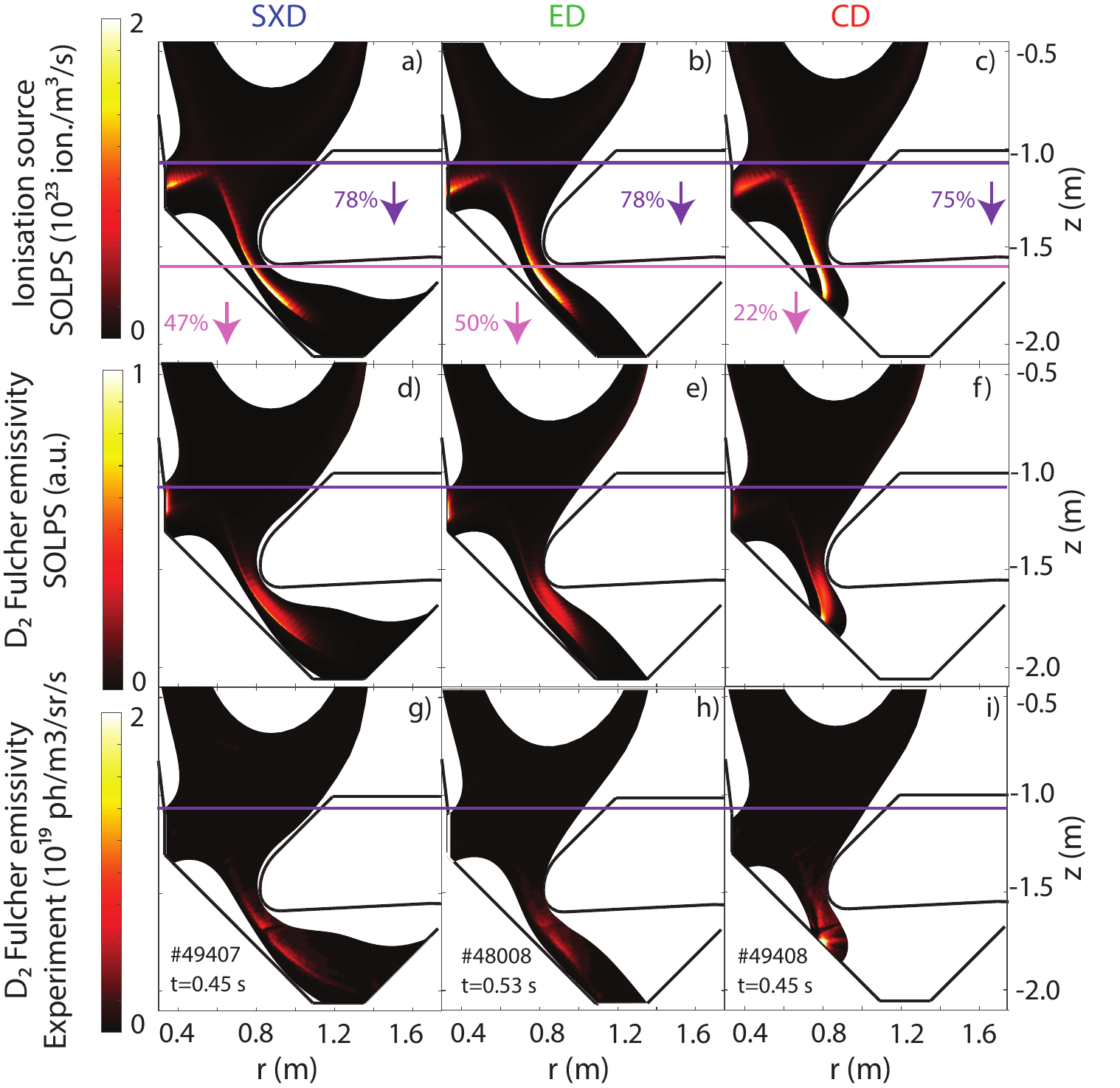}
	\caption{Overview of ion sources and $D_2$ Fulcher emission (as an ionisation location proxy \cite{Verhaegh2023,Wijkamp2023}) in the combined divertor and X-point region using simulations (a-f) for the SXD (a,d,g), ED (b,e,h) and CD (c,f,i) configurations, showing negligible ionisation upstream the X-point, consistent with experimental measurements (g-i). a-c) 2D ionisation source from SOLPS-ITER simulations (shown in Figure \ref{fig:FulcherSOLPSDetachEvolution}) with horizontal lines at $z=-1.6$ m (pink) and $z=1.07$ m (magenta), demarking the edge of the divertor spectroscopy view and X-point, respectively. The fraction of the ion source downstream these limits compared to the total ion source (outer leg only) are noted. d-f) Synthetic diagnostic for the $D_2$ Fulcher emissivity (arbitrary units) obtained from SOLPS-ITER simulations. g-i) Measured $D_2$ Fulcher emissivity (595-605 nm \cite{Wijkamp2023}) obtained from combined divertor imaging and X-point imaging inversions. The indicated time and discharges used are shown and are obtained from repeat discharges for the same core density as used in Figure \ref{fig:FulcherSOLPSDetachEvolution} Only emissivities obtained at the same $r,z$ corresponding to the simulation grids are shown. An inversion artefact is present near $r=0.85$ m, $z=-1.6$ m, where there is a gap in coverage between the X-point and divertor imaging systems.}
	\label{fig:xpt_imag_sim}
\end{figure}

This behaviour agrees with SOLPS-ITER predictions \cite{Moulton2023} of the CD, ED and SXD configurations (\autoref{fig:FulcherSOLPSDetachEvolution} a-c). The CD simulation is attached, whereas the SXD and ED simulations are detached. The radius of both the $D_2$ Fulcher emission front as well as the 5 eV contour, for the ED and SXD, remains near $r=0.95$ m. The $D_2$ Fulcher emission near the X-point region is also in agreement between experiments (Figure \ref{fig:xpt_imag_sim}), suggesting that the strong neutral trapping obtained in the simulations is consistent with the experiment. 

However, the simulations feature an attached inner target (Figure \ref{fig:xpt_imag_sim} (a-c)) with strong $D_2$ Fulcher emission near the inner strike point (Figure \ref{fig:xpt_imag_sim} (d-f)), which is not observed experimentally (Figure \ref{fig:xpt_imag_sim} (g-i)). This suggests the inner target power loading is negligible in the experiments and overestimated in SOLPS-ITER simulations. This requires further study including using multi-diagnostic, Bayesian, analysis techniques \cite{Greenhouse2024} to infer plasma parameters outside the divertor chamber.

A more detailed comparison between experiments and simulations is obtained by comparing their ion sources and sinks (\autoref{fig:FulcherSOLPSDetachEvolution} (g-i)) in the outer divertor chamber, indicating a quantitative agreement between experiments and simulations for the ion source and EIR. The MAR ion sinks are underestimated in the simulation in the detached region. This discrepancy is resolved when a corrected rate for molecular charge exchange is used in SOLPS-ITER, adopted from \cite{Verhaegh2023c} (\autoref{fig:FulcherSOLPSDetachEvolution} (g)).

\section{Discussion}
\label{ch:relevance}

Using the unique capabilities of MAST-U as a test bed for investigating novel divertor topologies reduces uncertainty in extrapolating current knowledge to reactor class devices by validating both reduced models and exhaust simulations: a crucial milestone for exploring ADCs as a reactor solution. There are, however, key differences between MAST-U and a reactor that must be addressed. 
\begin{enumerate}
	\item  Reactors will operate at higher power input and smaller heat flux widths than MAST-U. As a result, heat loads will become so large that impurity seeding will be required to lower target temperatures sufficiently to enable detachment. Understanding how combining total flux expansion, divertor neutral baffling, and poloidal leg length affects impurity-driven power dissipation requires further investigation.  Reactor-scale simulations have, however, demonstrated that ADCs enhance power dissipation, enabling operation with target heat fluxes below engineering limits with reduced impurity concentrations \cite{Xiang2021,Hudoba2023,Wigram2019}.
	
	Although higher power conditions impact neutral transport, plasma-molecular interactions are expected to play a key role in obtaining significant power, momentum and particle dissipation \emph{during (deep) detachment} according to exhaust simulations that incorporate long-legged, tightly baffled, divertors \cite{Verhaegh2023c,Hudoba2023,Osborne2024}. The critical role plasma-molecular chemistry plays in the ED and SXD illuminated discrepancies with simulations (\autoref{fig:FulcherSOLPSDetachEvolution} (h)) that have been reduced with improved rates for $D_2 + D^+ \rightarrow D_2^+ + D$. Extrapolating these improved rates to reactors with long-legged, tightly baffled, divertors shows they can make a critical impact on the reactor scale \cite{Verhaegh2023c}. 
	\item  Reactor-grade operation typically involves H(igh confinement)-mode operation. Current H-mode operation exhibits violent ELMs that result in extremely high heat loads. Although reactors will aim to minimize or suppress ELMs, it remains unclear whether ADCs can effectively mitigate the heat loads of residual (fast, ELM) transients to increase divertor lifetimes.
	\item Our MAST-U results show a balanced double-null configuration, with similar particle fluxes reaching the lower and upper outer divertors and negligible power reaching the inner target (Figure \ref{fig:xpt_imag_sim} \cite{Verhaegh2023c}). Achieving such up/down balance in reactors poses challenges due to a larger distance between the X-point and poloidal field coils \cite{Osawa2023} and smaller scrape-off-layer widths ($10-12$ mm for the current MAST-U experiments; $1-2$ mm for STEP \cite{Henderson2024} and $0.2-0.4$ mm for SPARC \cite{Kuang2020}). Although the exhaust benefits of double null may be limited according to reactor exhaust simulations \cite{AhoMantilla} and TCV experiments \cite{Fevrier2021}, double-null imbalances exacerbates inner target power loading and may necessitate solutions for spherical tokamak reactors, such as a proposed inner target X-Divertor geometry \cite{Hudoba2023,Henderson2024} requiring further experimental validation.
	\begin{itemize}
		\item One concern of outer target optimisation strategies is that the increased outer target connection length will exacerbate the inner target heat load in single null conditions \emph{in attached conditions} according to reduced models \cite{Maurizio2019}. This was in contrast to \emph{detached} DEMO Super-X divertor simulations, which indicated inner target heat loads were reduced with outer divertor optimisation \cite{Militello2021,Xiang2021}, requiring further study. 
		\item This reduced model (for \emph{attached} conditions) predicts an increase of inner target heat loads by 31 \% (ED) and 57 \% (SXD) compared to the CD. However, these increased inner target heat loads are reduced when accounting for outer target detachment. Using the ED and SXD outer divertor detachment front as a virtual target, the additional inner target heat loads are reduced below 15 \% compared to an outer target attached CD.
	\end{itemize}
\end{enumerate}

Future MAST-U experiments aim to address these key differences and advance towards more reactor-relevant scenarios. Planned upgrades include increased external heating from 4.4 MW to over 10 MW ($>2026$), enabling hotter, more attached divertor conditions. Cryopumping has been installed to reduce divertor neutral pressures and obtain hotter divertor conditions (2025). Advanced scenario development (2025) may enable enable single-to-double-null comparisons. Preliminary results suggest that the benefits of ADCs observed in this study persist in H-mode and may even buffer small ELMs, requiring further investigation \cite{Verhaegh2023b}. 

Overall, our results demonstrate that ADCs not only improve exhaust performance, enabling reduced upstream density and likely thus impurity concentration/core radiation in reactors \cite{Xiang2021}, but also improve core-edge compatibility when paired with strong baffling. This allows for  detached divertor operation without compromising core conditions: a major milestone towards proving the applicability of ADCs in reactors. However, any reactor design needs a compromise between engineering complexity and attractive operating regimes \cite{Militello2021,Kembleton2022}. The increased engineering complexity of some ADCs, such as the Super-X divertor, remains a significant consideration for reactor designs due to associated cost, space constraints and magnetic control tolerances \cite{Militello2021,Kembleton2022}. Our findings underscore that smaller modifications to divertor topology, such as transitioning from CD to ED, can achieve significant performance gains that are consistent with reduced model predictions and exhaust simulations. ADC design is therefore a continuum - an insight that has implications for reactor designs of DEMO, as well as more compact machines (SPARC, ARC, STEP) \cite{Henderson2024,Wigram2019,Kuang2020}, and paves the way for designs that optimise power exhaust and core-edge compatibility with reduced engineering and integration demands \cite{Militello2021,Kembleton2022}.

\section{Methods}\label{Methods}

\subsection{MAST Upgrade, Alternative Divertor Configurations and the Super-X Divertor}
\label{ch:MASTU}

MAST Upgrade is the UK's national fusion experiment tackling one of fusion energy's biggest challenges: plasma exhaust. It is a medium sized, small aspect ratio (i.e., spherical) tokamak (major radius: 0.9 m, minor radius: 0.6 m) operated by the United Kingdom Atomic Energy Authority \cite{Harrison2023,Morris2018}. It has a toroidal field of 0.8 T, its plasma current can reach up to 1 MA, and features one off-axis and one on-axis neutral beam  injector for external heating, of up to 2.2 MW each. TRANSP simulations are used to model the neutral beam absorption, required to estimate the power entering the scrape-off-layer (SOL). MAST-U features core Thomson scattering to obtain core electron density and temperature profiles, uses far-infrared-reflectometry (FIR) to obtain the line-averaged electron density and utilises the magnetic equilibrium reconstruction code EFIT++ to reconstruct the magnetic equilibria based on magnetic probe measurements. Its divertor is well diagnosed, featuring line-of-sight spectroscopy \cite{Verhaegh2023}, imaging diagnostics \cite{Wijkamp2023}, Langmuir probes \cite{Ryan2023}, as well as an imaging bolometry system at the X-point \cite{Federici2023}.

In this work, fuelling injection from the low field side of the core plasma is used to maintain L-mode conditions whilst controlling the core density in real time by adapting the low-field side main chamber fuelling rate \cite{Derks2024}. The advantage of using higher power L-mode conditions in this study is that the upstream density can be reduced compared to that in H-mode. This makes the plasma less detached and enables a wider range of upstream density scans to investigate the evolution during detachment \cite{Verhaegh2023c}. 

MAST-U features upper and lower divertor chambers, enabling double null diverted scenarios. The divertor chambers prevent neutral transport from the divertor to the core, providing neutral baffling and contributing to core-edge compatibility. The large divertor chamber, combined with various divertor coils \cite{Morris2018}, facilitates the integration of complex divertor shapes with strong neutral baffling. This enables studying the impact of divertor shaping on power exhaust while maintaining strong neutral baffling. 

With this shaping flexiblity, MAST-U can alter the poloidal flux expansion, connection length and total flux expansion. Poloidal flux expansion, $F_x = \frac{B_\theta^{u} B_\phi^t}{B_\theta^t B_\phi^u}$ \cite{Theiler2017}), is the ratio of the perpendicular flux surface spacing at the target and upstream, where $B_{\theta, \phi}^{u, t}$ are the poloidal ($\theta$) and toroidal ($\phi$) components of the magnetic field at upstream ($u$) and at the target ($t$), respectively. Increasing $F_x$ reduces the target heat loads (W/$m^2$) by spreading it over a larger surface. Increasing the connection length between the midplane and the divertor target ($L_\parallel$), provides a larger radiating volume and is expected to improve power exhaust \cite{Theiler2017}. Total flux expansion ($F_R = \frac{B_{xpt}}{B_t}$) increases the cross-sectional area of a flux tube, spreading the heat over a larger radius and lowering the target temperature  \cite{Theiler2017,Lipschultz2016}. The spherical nature of MAST-U enables varying total flux expansion over an unprecedented range, making it an ideal testbed for studying the impact of total flux expansion in a strongly baffled divertor.

\subsection{Divertor detachment, ion source/sink inferences and power balance}
\label{ch:Detachment_BaSPMI}

Power exhaust can be facilitated by plasma detachment, which is a state where simultaneous power, particle and momentum losses result in a simultaneous reduction of target particle fluxes and plasma target temperature.

Using novel spectroscopic techniques \cite{Verhaegh2021,Verhaegh2023} of hydrogen atomic Balmer line emission, the electron temperature, ion sources ($I_i$) and sinks ($I_r$) from plasma-atom and molecular interactions, as well as the hydrogenic radiative power losses and Molecular Activated Dissociation, can be inferred from the hydrogen Balmer line emission. 

Since the line-of-sight spectroscopy system has a set fan of views throughout the divertor leg (\autoref{fig:MagGeom}), spatial profiles of chordally integrated ion sources and sinks (part./$m^2$/s) along the divertor leg can be obtained (\autoref{fig:PartPowerBalSXD_ED_CD} d,e,f). During detachment, first the ionisation source detaches from the target ($T_e<3-5$ eV, inferred spectroscopically \cite{Verhaegh2023}) and ultimately Electron-Ion Recombination (EIR) starts to occur near the target ($T_e\approx0.2$ eV, $n_e \approx 2-4 \times 10^{19} m^{-3}$, according to spectroscopic inferences of the high-n ($n>9$) Balmer line spectra \cite{Verhaegh2023b}). By tracking the location of the downstream-end of the ionisation source ($(1.5 \pm 0.25) \times 10^{21} part/m^2/s$) and the upstream-end of the EIR sink ($(3 \pm 0.5) \times 10^{20} part/m^2/s$), the distance between the target and the ionisation front (defined as the detachment front) and EIR front (colder reference point of deeper detachment) can be obtained. These numbers are obtained as onset points based on the spatial profiles of ion sources and sinks presented in \autoref{fig:MagGeom} (d).

Combining spectroscopic inferences on ion sources and sinks with Langmuir probe measurements, information on both particle and power balance can be obtained. The total ion target flux ($I_t$ in part./s) is obtained by integrating the ion target flux $\Gamma_t$ (part/$m^2$/s) measured by Langmuir probes. From conservation of particles, the total ion target flux should equal the ion sources minus the ion sinks, in addition to any net ion inflow into the monitored system $I_u$, \autoref{eq:PartPowerBal}. Using particle balance, the total ion source ($I_i + I_u$) can be inferred (\autoref{eq:PartPowerBal}).

\begin{equation}
	I_t = I_i - I_r + I_u
	\label{eq:PartPowerBal}
\end{equation}

The target power loading can be inferred using a combination of spectroscopy and Langmuir probe measurements. To overcome limitations of estimating target temperatures using Langmuir probes in low temperature conditions \cite{Ryan2023}, spectroscopy from lines-of-sight closest to the target is used to infer a characteristic target electron temperature $T_t$. Using this temperature, the perpendicular plasma target power deposition can be estimated as $P_{\perp, target} = I_t (\gamma T_t + \epsilon)$ (in W), whereas the peak perpendicular heat flux can be estimated as  $q_{\perp, peak} = \Gamma_{t, peak} (\gamma T_t + \epsilon)$ (in W/$m^2$). A sheath transmission factor of $\gamma=7$ is assumed (valid for equal electron and ion temperatures) and both surface recombination and molecular re-association is accounted for in the potential energy $\epsilon = 13.6 + 2.2$ eV. 

Assuming that all volumetric power losses are purely due to hydrogenic radiation as well as dissociation, which is motivated by the observation that hydrogenic radiation estimates from spectroscopic analysis align with the measured total radiation in these conditions \cite{Verhaegh2023c}, the power into the divertor chamber can be estimated by summing $P_{\perp, target}$ and the inferred hydrogenic divertor radiative power loss. This ignores power transfer from the plasma to the neutral cloud through elastic collisions, which can become substantial in the SOLPS-ITER simulations shown (up to 15 \% of $P_{SOL}$ \cite{Osborne2024}), in qualitative agreement with the observed rotational $D_2$ rotational temperatures \cite{Osborne2024}. 

Although this does include surface recombination, it does not include target heat loads due to photons and neutral atoms. Including dissociation as a total loss channel implies assuming that the neutral atoms, after dissociation, are mostly lost to the side walls, rather than reaching the target (and hence do not contribute as target heating), which is consistent with findings in SOLPS-ITER simulations.

\subsection{Detachment Location Sensitivity (DLS) model}
\label{ch:DLS_model}

The Detachment Location Sensitivity (DLS) analytical model \cite{Lipschultz2016,Cowley2022,Myatra2023a} can model the impact of the magnetic divertor geometry on the detachment threshold in terms of changes to a control parameter $C \propto \frac{n_u \sqrt{f_z}}{q_\parallel^{5/7}}$, which depends on upstream density $n_u$, impurity fraction $f_z$ and parallel heat flux $q_\parallel$. The detachment onset $C_t$ is proportional to a term that only depends on the magnetic geometry \cite{Myatra2023a}, as shown in \autoref{eq:DLS_onset}. In here, $B$ is the total magnetic field, $\xi$ is the coordinate representing the volume of the flux tube between the target and a given position along the divertor leg, scaled by a reference area $\propto 1/B_{xpt}$. 

\begin{equation}
	C_t \propto \frac{B_t}{B_{xpt}^{3/7}} \times (\int_t^{xpt} B^2 (\xi) d\xi + \int_{xpt}^u B^2 (\xi) (\frac{L-\xi}{L-\xi_{xpt}}) d\xi)^{-2/7} 
	\label{eq:DLS_onset}
\end{equation}

The advantage of this formulation is that it considers the full magnetic field dependency numerically, rather than approximating the field variation as linear with $\xi$. Under those approximations, \autoref{eq:DLS_onset}: $C_t \propto \frac{B_t}{B_{xpt}} (\frac{B_{xpt}}{<B>})^{2/7} \frac{1}{L_\parallel^{2/7}}$ \cite{Cowley2022}. The DLS model thus predicts that detachment onset is facilitated by increased connection length ($L_\parallel$) and increased total flux expansion $\frac{B_t}{B_{xpt}}$. We find negligible differences between this approximate form and the full numerical calculation for the MAST-U shapes reported in this work. The DLS model is applied to a flux tube that is 0.5 mm outwards of the separatrix into the SOL, to avoid numerical errors. Instead of finding the detachment onset, where a 'thermal front' (i.e., detachment front) leaves the target, equation \ref{eq:DLS_onset} can also be applied to any position of the detachment front along the leg by changing the target to a different location. This implies that the front location only depends on the magnetic geometry upstream of the front. By monitoring how quickly the front position changes along the divertor leg, the DLS can make predictions on detachment front sensitivity.

The DLS model formally assumes that all power is dissipated by impurity radiation and that the radiating specie has a constant concentration in the radiating region. This is likely not the case in the MAST-U \emph{divertor chamber} where the radiative power losses are dominated by hydrogenic interactions \cite{Verhaegh2023} consistent with SOLPS-ITER modelling predictions \cite{Moulton2023}, although impurity radiation could be significant upstream of the divertor chamber entrance \cite{Verhaegh2023,Verhaegh2023c}. However, the impact of the divertor topology on the detachment onset appears to be more generally applicable outside of impurity radiation dominant conditions. 

Assuming the impurity fraction is constant and that the upstream electron density and heat flux are proportional to, and fully determined by, $f_{GW}$ and $P_{SOL}$, the detachment threshold is expected to be dependent on $f_{GW}$ and $P_{SOL}$. There is a small variation in $P_{SOL}$ during the experiment (increased NBI power absorption at higher densities), which is accounted for in our predictions. In our detached conditions, $\lambda_q$ cannot be monitored from target measurements. However, scaling laws on MAST (attached, open, conventional divertor) did show an increase in $\lambda_q$ at higher $f_{GW}$ \cite{Harrison2013} and this scaling law dependency is accounted for in our predictions. Not accounting for this predicted change in $\lambda_q$ and for $P_{SOL}$ changes only has a secondary impact on the predicted $f_{GW}$ for the various detachment threshold and does not impact any of the conclusions in this work.

\subsection{Exhaust simulations - SOLPS-ITER}
\label{ch:SOLPS}

Reduced models, such as the Detachment Location Sensitivity (DLS) model, are useful for building a physics understanding. However, the divertor behaviour is highly complex: it is a 2D/3D phenomena that involves interactions between the plasma with neutral atoms and molecules. SOLPS-ITER is a state-of-the-art code suite for advanced power exhaust modelling \cite{Wiesen2015}. It combines a fluid code (B2.5) with a Monte Carlo neutral code that tracks the neutrals and incorporates several atomic and molecular databases (Eirene) \cite{Wiesen2015}.

Interpretive SOLPS-ITER simulations have been performed using a baseline Super-X SOLPS-ITER setup that has been matched against Ohmic experimental data \cite{Moulton2023}. These simulations have been extrapolated to the higher power experiments presented here, using corresponding experimental magnetic equilibria for the Super-X, Elongated and Conventional divertors, fuelling location, and $P_{SOL}$. The fuelling rate has been tuned in order to match the experimentally measured upstream electron densities.

For the SOLPS-ITER simulation setup (Super-X divertor only) with corrected rates ('Sim. Corr. Rate (SXD), \autoref{fig:FulcherSOLPSDetachEvolution}g), the molecular charge exchange rate ($D_2 + D^+ \rightarrow D_2^+ + D$) has been replaced with a newly calculated rate that has been specifically computed for deuterium \cite{Verhaegh2023c}. This rate uses ab inito calculated cross-sections from \cite{Ichihara2000} and combines them with a collisional-radiative model calculation (to compute the vibrational distribution of $H_2$ molecules) utilising the same data (apart from molecular charge exchange) that is used in Eirene. Furthermore, it accounts for isotope mass differences correctly \cite{Verhaegh2023c,Verhaegh2023a}, as well as for the differences in vibrational energy levels between the different isotopes \cite{Verhaegh2023c,Ichihara2000}. This improved rate is significantly increased at low temperatures ($T<2$ eV), particularly for the heavier isotopes, compared to the default rate used in Eirene \cite{Verhaegh2023a,Verhaegh2023c}.








	
	
	

\section{Data availability}

The data that support these studies are openly available at: \url{https://doi.org/10.14468/9m0q-kc26}. Code and software used to generate the results in this paper are referenced in the metadata of this DOI. To obtain further information on the data and models underlying this paper please contact publicationsmanager@ukaea.uk. 

\section{Acknowledgements}

This work has received support from EPSRC Grants EP/T012250/1, EP/N023846/1 and EP/W006839/1. This work is supported by US Department of Energy, Office of Fusion Energy Sciences under the Spherical Tokamak program, contract DE-AC05-00OR22725. This work has been carried out within the framework of the EUROfusion Consortium, partially funded by the European Union via the Euratom Research and Training Programme (Grant Agreement No 101052200 — EUROfusion). The Swiss contribution to this work has been funded by the Swiss State Secretariat for Education, Research and Innovation (SERI). Views and opinions expressed are however those of the author(s) only and do not necessarily reflect those of the European Union, the European Commission or SERI. Neither the European Union nor the European Commission nor SERI can be held responsible for them.

\section{Author contributions (CReDiT)}

Grouped co-authors indicate equal contributions.

\begin{itemize}
	\item \textbf{Kevin Verhaegh:} Conceptualisation, Methodology, Software, Validation, Formal Analysis, Investigation, Data curation, Writing - Original Draft, Writing - Review \& Editing, Visualisation, Supervision, Project Administration, Funding acquisition.
	\item \textbf{James Harrison:} Conceptualisation, Software, Resources, Writing - Review \& Editing, Supervision, Funding Acquisition. \textbf{David Moulton:} Conceptualisation, Formal Analysis, Investigation, Writing - Review \& Editing.
	\item \textbf{Bruce Lipschultz:} Conceptualisation, Investigation, Supervision, Resources, Project administration. \textbf{Nicola Lonigro:} Investigation, Formal Analysis, Validation, Visualisation. \textbf{Nick Osborne:} Investigation, Validation. \textbf{Peter Ryan:} Formal Analysis, Validation, Visualisation. \textbf{Christian Theiler:} Conceptualisation, Investigation, Writing - Review \& Editing, Project administration, Supervision, Funding acquisition. \textbf{Tijs Wijkamp:} Investigation, Formal analysis, Software, Visualisation, Writing - Review \& Editing.
	\item \textbf{Dominik Brida:} Investigation, Project administration, Supervision, Funding acquisition. \textbf{Cyd Cowley:} Validation, Methodology. \textbf{Gijs Derks:} Investigation. \textbf{Rhys Doyle:} Investigation. \textbf{Fabio Federici:} Investigation, Writing - Review \& Editing. \textbf{Bob Kool:} Investigation. 
	\item \textbf{Olivier F\'{e}vrier} Project administration, Funding acquisition. \textbf{Antti Hakola:} Conceptualisation, Project administration, Funding acquisition. \textbf{Stuart Henderson}: Investigation. \textbf{Holger Reimerdes:} Project administration, Funding acquisition. \textbf{Andrew Thornton:} Investigation. \textbf{Nicola Vianello:} Project administration, Funding acquisition, Supervision. \textbf{Marco Wischmeier:} Project administration, Funding acquisition, Supervision. \textbf{Lingyan Xiang:} Formal analysis.
\end{itemize} 

\section{References}

\bibliographystyle{iopart-num}
\bibliography{all_bib.bib}
\end{document}